\documentclass{pasj00}
\draft
\SetRunningHead{\textsc{IKEYA} and \textsc{MATSUMOTO}}{Stability Property of Numerical Cherenkov Radiation}
\Received{2014 December 2}
\Accepted{2015 May 27}
\begin{document}
\title{Stability Property of Numerical Cherenkov Radiation and its Application to Relativistic Shock Simulations}
\author{Naoki \textsc{Ikeya} and Yosuke \textsc{Matsumoto}}
\affil{Department of Physics, Chiba University, 1-33 Yayoi, Inage-ku, Chiba, 263-8522}
\email{ikeyank@astro.s.chiba-u.ac.jp, ymatumot@chiba-u.jp}


\maketitle

\begin{abstract}
We studied the stability property of numerical Cherenkov radiation in relativistic plasma flows employing particle-in-cell simulations. Using the implicit finite-difference time-domain method to solve Maxwell equations, we found that nonphysical instability was greatly inhibited with a CFL number of 1.0. The present result contrasts with recently reported results (\cite{Vay_11}; \cite{Godfrey_Vay_13}; \cite{Xu_13})  in which magical CFL numbers in the range 0.5--0.7 were obtained with explicit field solvers. In addition, we found employing higher-order shape functions and an optimal implicitness factor further suppressed long-wavelength modes of the instability. The findings allowed the examination of the long-term evolution of a relativistic collisionless shock without the generation of nonphysical wave excitations in the upstream. This achievement will allow us to investigate particle accelerations in relativistic shocks associated with, for example, gamma-ray bursts.
\end{abstract}

\section{Introduction}
Particle-in-cell (PIC) simulations have been used to study the plasma dynamics of laboratory, space, and astrophysical phenomena. In particular, PIC simulations have been powerful tools in order to investigate particle accelerations associated with explosive phenomena in astrophysical objects, such as supernova remnant shocks (e.g., \cite{Matsumoto_13}) and gamma-ray bursts (e.g., \cite{Sironi_Spitkovsky_11}). In PIC simulations, the finite-difference time-domain (FDTD) method has been employed as a standard numerical solver for the Maxwell equations. The FDTD method is simple and flexible but its use has been known to result in the phase speed of the electromagnetic wave being numerically less than the speed of light in high wave number regions because of the finite size of spatial cells and time steps. In relativistic plasma flows, this numerical dispersion induces a nonphysical numerical instability that is now known as numerical Cherenkov radiation \citep{Godfrey_74}. This is one of the critical issues in examining relativistic collisionless shocks employing multidimensional PIC simulations with the FDTD method. Several methods have been developed to suppress the numerical instability. One method is to apply digital filtering to waves in high wave number regions. While this approach has been widely applied in relativistic plasma simulations (\cite{Greenwood}; \cite{Spitkovsky_08b}; \cite{Martins}; \cite{Vay_11}), the filters may induce numerical damping of physical waves. This becomes problematic particularly when examining particle accelerations in collisionless shocks. Another options are the family of spectral methods (\cite{Haber}; \cite{Lin}; \cite{Dawson}) , in which the Maxwell equations are solved in Fourier space. Thereby, the numerical Cherenkov radiation are eliminated by a cutoff filter directly applied in large wave number regions (\cite{Nagata}; \cite{Yu}). However, since the spectral methods are based on the fast Fourier transformation, PIC simulations employing these methods are still under investigation in terms of their computational cost when conducted on distributed-memory systems (\cite{Vay_13}).

It has recently been reported that careful choices of the CFL number greatly inhibited growth of the numerical Cherenkov instability \citep{Vay_11}. For specific PIC simulation algorithms employing explicit FDTD field solvers and the density decomposition method for the current deposit \citep{Esirkepov_01}, a dispersion relation of the numerical Cherenkov instability has been derived (\cite{Godfrey_Vay_13}; \cite{Xu_13}). These results indicate that magical CFL numbers, which minimize the imaginary part of the dispersion relation, are approximately 0.5--0.7 depending on the PIC algorithm. In this paper, we examine this stability property of the numerical Cherenkov instability by means of two-dimensional PIC simulations. We used a PIC simulation code package, pCANS, which employs momentum-conserving field interpolation \citep{Birdsall}, the density decomposition method, and the implicit FDTD method for the Maxwell equations. Using the present implicit field solver, we expect to obtain stability properties different from the previous results obtained with explicit field solvers.

The present paper is organized as follows. Section 2 describes the dispersion relation of the implicit field solver in the Fourier space. Section 3 presents numerical tests for various CFL numbers, orders of the shape function and implicitness factors. Section 4 proposes optimal choices of these numerical parameters and then applies the choices to relativistic collisionless shock simulations. Section 5 summarizes results in view of the relativistic shock simulations.

\section{Numerical Cherenkov radiation}
To investigate numerical Cherenkov instability, we derived a numerical dispersion relation of the electromagnetic waves with an implicit field solver employed in pCANS code. The electromagnetic fields are advanced according to the Maxwell equations
\begin{eqnarray}
\frac{\partial \boldsymbol{B}}{\partial t}=&-c\nabla \times \boldsymbol{E},
\label{eq:maxwell1}\\
\frac{\partial \boldsymbol{E}}{\partial t}=& \:\:c\nabla \times \boldsymbol{B}&-4\pi \boldsymbol{J},
\label{eq:maxwell2}
\end{eqnarray}
where $\boldsymbol{B}$, $\boldsymbol{E}$, $\boldsymbol{J}$ and $c$ denote the magnetic field, electric field, current density, and speed of light, respectively.

\begin{figure}
 \begin{center}
   \FigureFile(80mm,80mm){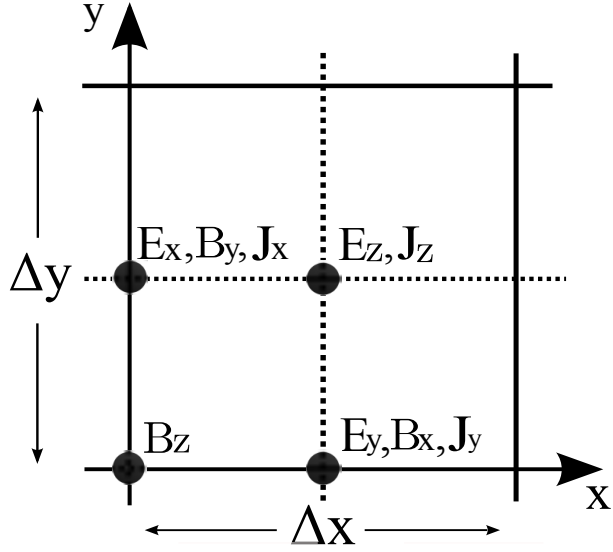}
 \end{center}
\caption{Two-dimensional spatial arrangement of the electromagnetic fields and current density in the FDTD method.}\label{fig:Yee}
\end{figure}

In the explicit FDTD method, the electric and magnetic fields and the current density are defined on a staggered grid system as depicted in Figure \ref{fig:Yee} and at different time steps as
\begin{eqnarray}
&&\frac{\boldsymbol{B}^{n+1/2}-\boldsymbol{B}^{n-1/2}}{\Delta t}=-c\nabla \times \boldsymbol{E}^{n}, 
\label{eq:maxwell_FDTD1}\\
&&\frac{\boldsymbol{E}^{n+1}-\boldsymbol{E}^{n}} {\Delta t}=c\nabla \times \boldsymbol{B}^{n+1/2}-4\pi\boldsymbol{J}^{n+1/2}.
\label{eq:maxwell_FDTD2}
\end{eqnarray}
In the present implicit solver, the Maxwell equations are discretized in time as
\begin{eqnarray}
&&\frac{\boldsymbol{B}^{n+1}-\boldsymbol{B}^n}{\Delta t}=-c\nabla \times(\theta \:\boldsymbol{E}^{n+1} + (1 - \theta) \boldsymbol{E}^n),
\label{eq:maxwell_diff1}\\
&&\frac{\boldsymbol{E}^{n+1}-\boldsymbol{E}^n}{\Delta t}=c\nabla \times(\theta \:\boldsymbol{B}^{n+1} + (1 - \theta) \boldsymbol{B}^n)-4\pi\boldsymbol{J}^{n+1/2},
\label{eq:maxwell_diff2}
\end{eqnarray}
where $\theta$ is an implicitness factor that ranges between 0.5 and 1.0. (Note that $\theta$ = 0.5 and 1.0 correspond to the Crank--Nicolson scheme and the backward Euler scheme, respectively.) 

Letting $\delta \boldsymbol{B} = \boldsymbol{B}^{n+1}-\boldsymbol{B}^n$, it follows from Eqs. (\ref{eq:maxwell_diff1})  and (\ref{eq:maxwell_diff2}) that
\begin{eqnarray}
\left[1-\left( \theta c \Delta t \right)^2 \nabla^2 \right] {\delta \boldsymbol{B}} = \theta \left(c \Delta t \right)^2 \left( \nabla^2 {\boldsymbol{B}}^n + \frac{4\pi}{c} \nabla \times {\boldsymbol{J}}^{n+1/2} \right) -c \Delta t \nabla \times {\boldsymbol{E}}^n.
\label{eq:conjugate}
\end{eqnarray}
Discretizing in space with boundary conditions, the solution for $\delta \boldsymbol{B}$ in Eq. (\ref{eq:conjugate}) can be obtained employing the conjugate gradient method \citep{Hoshino_13}. Although the computational cost of this implicit method is generally greater than that of the explicit FDTD method, the total calculation cost is not that much greater when the implicit method is coupled with the PIC algorithm, which dominates the overall computational cost. 

The electric and magnetic fields can be expressed as the plane waves with the time index $n$ and the grid indexes $l$ and $m$. 
\begin{eqnarray}
\boldsymbol{B}= \boldsymbol{B_0}\exp\{i( -\omega n \Delta t + k_x l \Delta x + k_y m\Delta y)\}\label{eq:plane_wave_B},\\
\boldsymbol{E}= \boldsymbol{E_0}\exp\{i( -\omega n \Delta t + k_x l \Delta x + k_y m\Delta y)\},
\label{eq:plane_wave_E}
\end{eqnarray}
Substituting Eqs. (\ref{eq:plane_wave_B}) and (\ref{eq:plane_wave_E}) into Eqs. (\ref{eq:maxwell_diff1}) and (\ref{eq:maxwell_diff2}), one obtains a numerical dispersion relation of the electromagnetic wave in vacuum ($\boldsymbol{J} = 0$) for the implicit FDTD method as
\begin{eqnarray}
\Biggl[\frac{(\mathrm{e}^{i \omega \Delta t}-1)}{\theta(\mathrm{e}^{i \omega \Delta t}-1)+1}\Biggl]^2
&&+4\biggl(c\frac{\Delta t}{\Delta x}\sin{\Bigl(\frac{k_x\Delta x}{2}\Bigr)}\biggr)^2+4\biggl(c\frac{\Delta t}{\Delta y}\sin{\Bigl(\frac{k_y\Delta y}{2}\Bigr)}\biggr)^2 = 0. 
\label{eq:dispersion_relation_i_a}
\end{eqnarray}
The dispersion relation reads as 
\begin{eqnarray}
&&\Re\bigl[\omega\Delta t\bigr]=2\arctan\Biggl[\frac{1}{4}\frac{\sqrt{\frac{1}{{S_x}^2+{S_y}^2}}}
{(\theta-\frac{1}{2})^2+\frac{1}{4}{\frac{1}{{S_x}^2+{S_y}^2}}}
\Biggr],
\label{eq:dispersion_real}\\
&&\Im\bigl[\omega\Delta t\bigr]=2\arctan\Biggl[\frac{1}{2}\frac{(\theta - \frac{1}{2})}{(\theta-\frac{1}{2})^2+\frac{1}{4}{\frac{1}{{S_x}^2+{S_y}^2}}}\Biggr],
\label{eq:dispersion_imp}
\end{eqnarray}
where
\begin{eqnarray}
S_x = c \frac{\Delta t}{\Delta x} \sin{(\frac{k_x \Delta x}{2})},\\
S_y = c \frac{\Delta t}{\Delta y} \sin{(\frac{k_y \Delta y}{2})},
\end{eqnarray}
and $\Re$ and $\Im$ are real and imaginary parts of the dispersion relation, respectively. Since the imaginary part is always positive provided $\theta \geq 0.5$, the implicit FDTD method damps high-frequency waves except in the case that $\theta=0.5$.

\begin{figure}
 \begin{center}
   \FigureFile(80mm,80mm){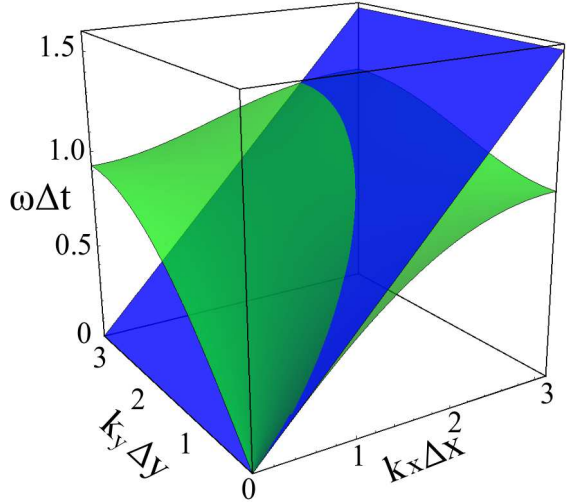}
 \end{center}
\caption{Green plane: visualization of the numerical dispersion relation of Eq. (\ref{eq:dispersion_relation_i_b}). Blue plane: the dispersion relation of the entropy mode moving at a speed $V = 0.9999c$.}\label{fig:Dispersion}
\end{figure}

In the case that $\theta$ = 0.5, the dispersion relation of Eq. (\ref{eq:dispersion_relation_i_a}) becomes
\begin{eqnarray}
\tan^2\Bigl(\frac{\omega \Delta t}{2}\Bigr) - {S_x}^2 - {S_y}^2 = 0.
\label{eq:dispersion_relation_i_b}
\end{eqnarray}
The dispersion relation of the explicit FDTD method is slightly different from the implicit one and is expressed as 
\begin{eqnarray}
\sin^2\Bigl(\frac{\omega \Delta t}{2}\Bigr) -{S_x}^2 - {S_y}^2 = 0.
\label{eq:dispersion_relation_e}
\end{eqnarray}
We present a numerical solution of Eq. (\ref{eq:dispersion_relation_i_b}) with $c\frac{\Delta t}{\Delta x} = c\frac{\Delta t}{\Delta y} = 0.5$ for the electromagnetic wave in Figure \ref{fig:Dispersion} (green plane), which has a dispersive aspect in high wave number regions. The short-wavelength modes thus propagate at a speed lower than the actual speed of light.

Here we suppose that plasma travels in the $x$-direction with bulk velocity $V = 0.9999c$. The dispersion relation of the entropy mode associated with movement of the plasma is $\omega = V k_x$ (blue plane in Figure \ref{fig:Dispersion}). Numerical Cherenkov radiation occurs if plasma flow overtakes the electromagnetic waves. This happens in wave number regions where the entropy mode crosses the electromagnetic wave. These intersections are obtained by substituting $\omega = V k_x$ into Eq. (\ref{eq:dispersion_relation_i_b}):
\begin{equation}
k_y\Delta y = 2\arcsin{\Biggl\{ \pm\frac{\Delta y}{c\Delta t}\sqrt{\tan^2\Bigl(\frac{Vk_x\Delta t}{2}\Bigr) - {S_x}^2} \Biggr\}}.
\label{eq:theory_line}
\end{equation}
We expect the numerical Cherenkov radiation to be destabilized at the wave numbers regulated by Eq. (\ref{eq:theory_line}). The relation is verified in the following two-dimensional PIC simulations.

\section{PIC simulations of numerical Cherenkov radiation}

We examined the stability of numerical Cherenkov radiation in two-dimensional PIC simulations. We used the PIC simulation package pCANS, which employs momentum-conserving interpolation for the fields, the density decomposition scheme for the current deposit, and the implicit field solver.

We conducted PIC simulations in a two-dimensional $x-y$ plane with a periodic boundary condition in each direction. We initially set a plasma flow in the $x$-direction with relativistic speed $V = 0.9999c$, which corresponds to a bulk Lorentz factor of $\Gamma = 1/\sqrt{1-(V/c)^2} = 100$. The thermal velocity in the rest frame of the flow was $10\%$ of the speed of light. The electric and magnetic fields were initially set to zero over the entire simulation box. We used 100 particles per cell for positrons and electrons in the simulation domain with $128 \times 128$ cells. Each cell size was equal to the Debye length. A second-order shape function for the particle and implicitness factor of $\theta$ = 0.501 were used in the following tests unless otherwise stated.

\subsection{Numerical Cherenkov radiation}

\begin{figure}
 \begin{center}
   \FigureFile(160mm,80mm){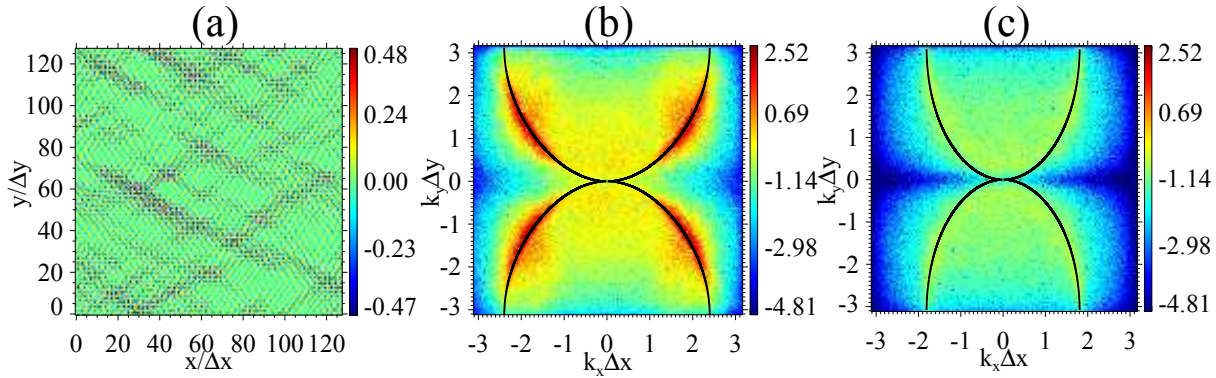}
 \end{center}
\caption{The $z$-component of the magnetic field $B_z$ in real space with a CFL number of (a) 0.5 and in Fourier space with CFL numbers of (b) 0.5 and (c) 1.0 at $t$ = 230 ($\Delta x /c$). The color expressed the $z$-component on a logarithmic scale. Eq. (\ref{eq:theory_line}) is plotted as the solid line in (b) and (c) and indicates the wave number regions of the radiation. The magnetic field was normalized by $\sqrt{4 \pi n_e m_e \Gamma c^2}$.}\label{fig:bz}
\end{figure}

We first present results from the simulation run with the CFL number $\sigma = c \frac{\Delta t}{\Delta x}$ = 0.5. Waves with large amplitudes were excited on grid scales (Figure \ref{fig:bz} (a)). The maximum amplitude reached a high level corresponding to 23\% of the kinetic energy of the plasma flow. Since the simulation was conducted for homogeneous streaming plasma, the excited strong waves should be nonphysical. The waves were excited in characteristic regions of the wave number space as shown in Figure \ref{fig:bz} (b); these are indeed the intersections indicated by Eq. (\ref{eq:theory_line}) (solid line in Figure \ref{fig:bz} (b)) and typical signatures of numerical Cherenkov radiation. 

\subsection{Stability dependence on the CFL number}

To investigate the stability properties of the Cherenkov instability, we conducted simulation runs with various CFL numbers ranging from 0.4 to 1.0. Note that the CFL number of 1.0 still gives a numerically stable solution owing to the use of the present implicit FDTD scheme. The total energy is conserved within $1\%$ error in all simulation runs.

\begin{figure}
 \begin{center}
  \FigureFile(160mm,80mm){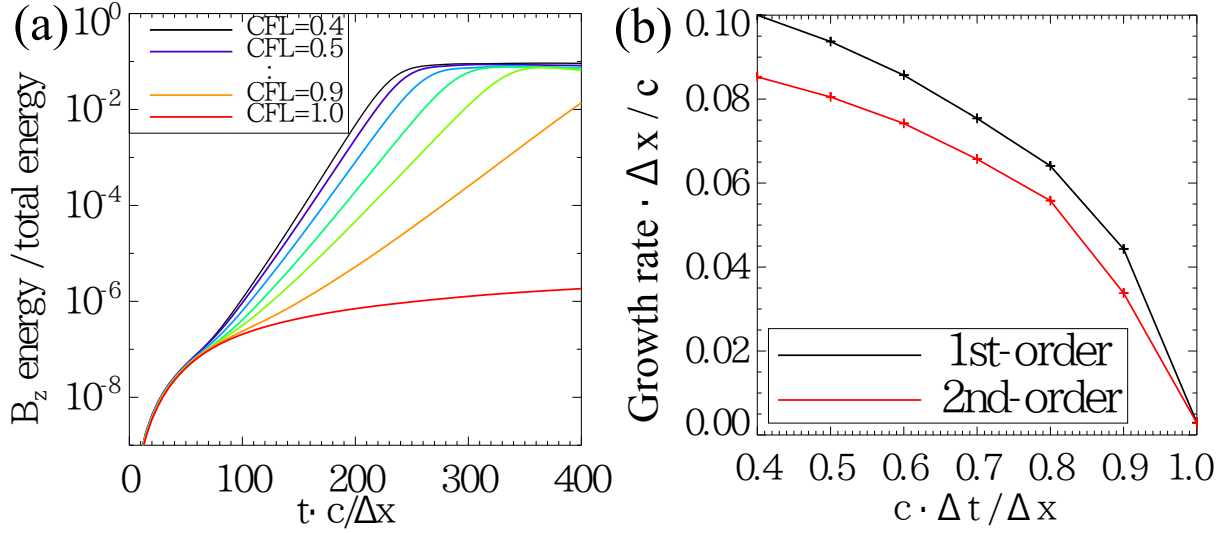} 
 \end{center}
\caption{(a) Temporal evolutions of the magnetic energy for various CFL numbers ranging from 0.4 to 1.0. (b) Growth rate of the numerical instability as a function of the CFL number for the first-order (black) and second-order (red) shape functions.}\label{fig:CFL}
\end{figure}

Figure \ref{fig:CFL} (a) shows the temporal evolution of the magnetic energy for various CFL numbers. Undesired growth of the magnetic energy was found in all simulation runs expect in the case of $\sigma$ = 1.0 (red line). Although the linear growth rate decreased as the CFL number approached 1.0, it saturated at 10\% of the total energy in the course of time. In contrast, the growth of the magnetic energy when $\sigma$ = 1.0 was within the noise level until $t$ = 400 ($\Delta x / c$). The growth rate of the instability as a function of $\sigma$ is shown in Figure \ref{fig:CFL} (b). The numerical instability was remarkably inhibited when $\sigma$ = 1.0. Figure \ref{fig:bz} (c) shows the result when $\sigma$ = 1.0 in the same format as Figure \ref{fig:bz} (b). 
The numerical Cherenkov instability was clearly inhibited when compared with the case of $\sigma$ = 0.5 in Figure \ref{fig:bz} (b).

\subsection{Slow-growth mode of the numerical Cherenkov instability}

\begin{figure}
 \begin{center}
   \FigureFile(160mm,80mm){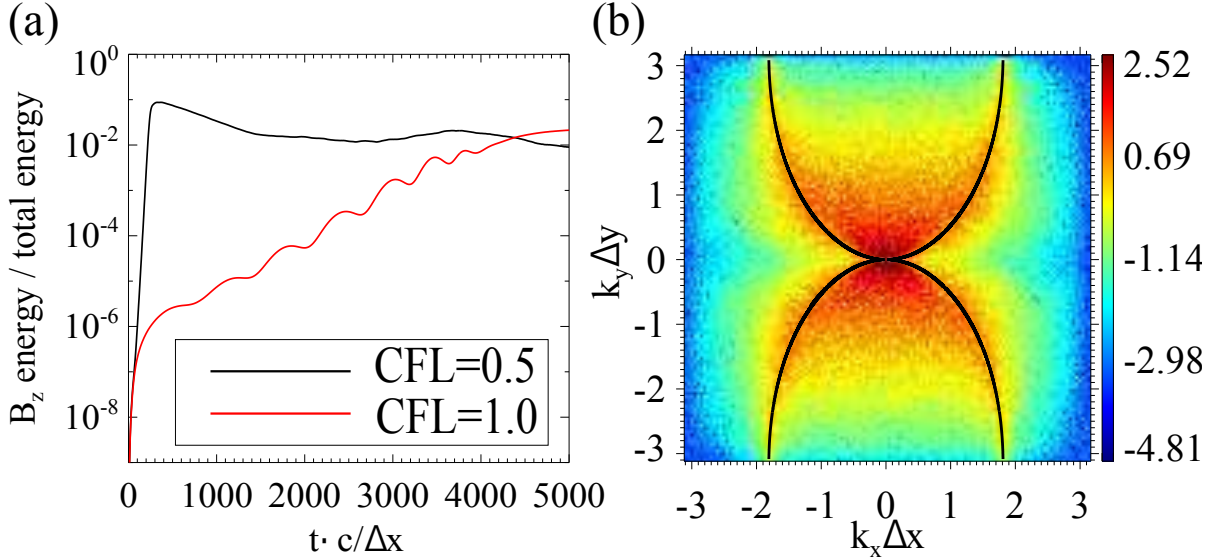}
 \end{center}
\caption{(a) Time evolution of the magnetic energy for CFL numbers of 0.5 (black) and 1.0 (red). (b) The $z$-component of the magnetic field $B_z$ for a CFL number of 1.0 at $t$ = 5000 ($ \Delta x /c$). The solid line shows the intersections obtained from Eq. (\ref{eq:theory_line}).}\label{fig:non_res}
\end{figure}

\begin{figure}
 \begin{center}
   \FigureFile(160mm,80mm){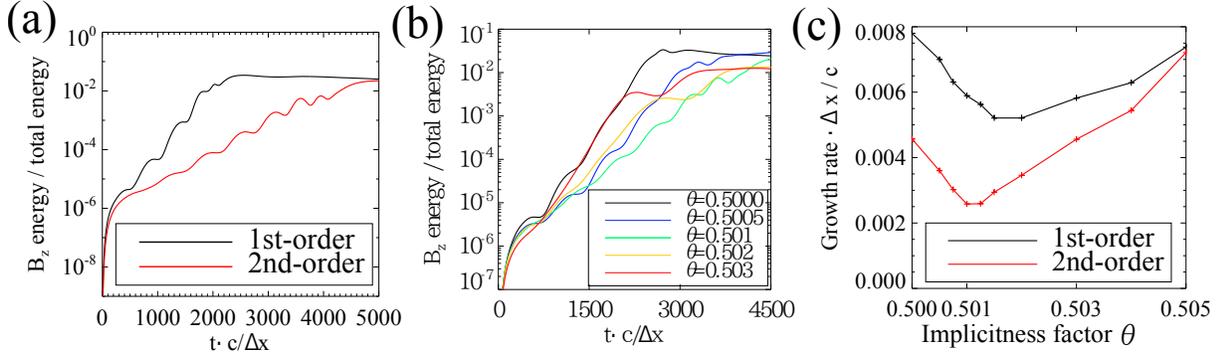}
 \end{center}
\caption{(a) Time evolutions of the magnetic energy for the first-order (black) and second-order (red) shape functions. (b) Time evolutions of the magnetic energy for various implicitness factors. (c) Growth rates as a function of the implicitness factor for the first-order (black) and second-order (red) shape functions.}\label{fig:non_depend}
\end{figure}

Following the simulation run with $\sigma$ = 1.0 until t = 5000 ($\Delta x / c$), we found another type of numerical Cherenkov instability that grew slowly and saturated at 10\% of the total energy (Figure \ref{fig:non_res} (a)). This was definitely a different mode from the mode previously discussed as it destabilized in lower wave number regions (Figure \ref{fig:non_res} (b)). The instability appeared in low wave number regions where physical waves of interest generally coexist. This is problematic in relativistic PIC simulations because digital filtering cannot be applied to long-wavelength modes. In this subsection, we present stability properties of this slowly growing mode using different shape functions and implicitness factors.

Figure \ref{fig:non_depend} (a) shows the temporal evolutions of the magnetic energy for different orders of the shape function on a much longer time scale (t $\sim$ 5000 ($\Delta x/c$)) than that in Figure \ref{fig:CFL} (a). We found that employing higher-order shape functions notably reduces the growth rate of this type of the instability. This property contrasts with that for the fast-growing mode in Figure \ref{fig:CFL} (b), in which only a slight difference was observed. 

We also found an optimal choice of the implicitness factor in terms of the slowly growing instability. Figure \ref{fig:non_depend} (b) shows the history of the magnetic energy for different implicitness factors with the second-order shape function. The growth rate decreased as $\theta$ increased from 0.5 to 0.501 and then turned to increase as $\theta$ increased from 0.501 to 0.503. This stability property for different orders of the shape function is summarized in Figure \ref{fig:non_depend} (c). We found optimal implicitness factors that minimized the growth rate of the slowly growing instability; these were $\theta_1$ = 0.502--0.503 and $\theta_2$ = 0.501 for the first- and second-order shape functions, respectively.

\section{Application to relativistic collisionless shock simulations}

In this section, we present two-dimensional PIC simulations of a relativistic shock with different CFL numbers. We adopted the second-order shape function and the implicitness factor of $\theta$ = 0.501 in the following comparisons.

Simulations were conducted on $N_x \times N_y = 5650 \times 990$ cells. Positrons and electrons were continuously injected from the boundary on the left-hand side of the simulation domain toward the $+x$ direction at 99.99\% of the speed of light, which corresponds to a bulk Lorentz factor of $\Gamma = 100$. The thermal velocity in the flow frame was 5\% of the speed of light. Fifty particles of each species per cell were used initially. Space and time were normalized by the electron skin depth $c / \omega_{\mathrm{pe}}$ and ${\omega_{\mathrm{pe}}}^{-1}$, respectively, where $\omega_{\mathrm{pe}} = \sqrt{4 \pi n_0 e^2 / m_e \Gamma}$ is the electron plasma frequency in the flow frame. The injected particles were reflected at the boundary on the right-hand side. Thus, evolutions were followed in the downstream rest frame in which shock waves propagated in the $-x$ direction. Interactions between the plasma traveling in the $+x$ direction and the reflected particles generated magnetic fields via the Weibel instability, which in turn decelerated the injected particles. A so-called Weibel-mediated shock formed in the present simulations (\cite{Kato_07}; \cite{Spitkovsky_08b}). The simulations were performed on a workstation using Intel Xeon processors and 230 GB of physical memory, and were parallelized via domain decomposition in the transverse direction into 32 processes. 

\begin{figure}
 \begin{center}
 \FigureFile(150mm,200mm){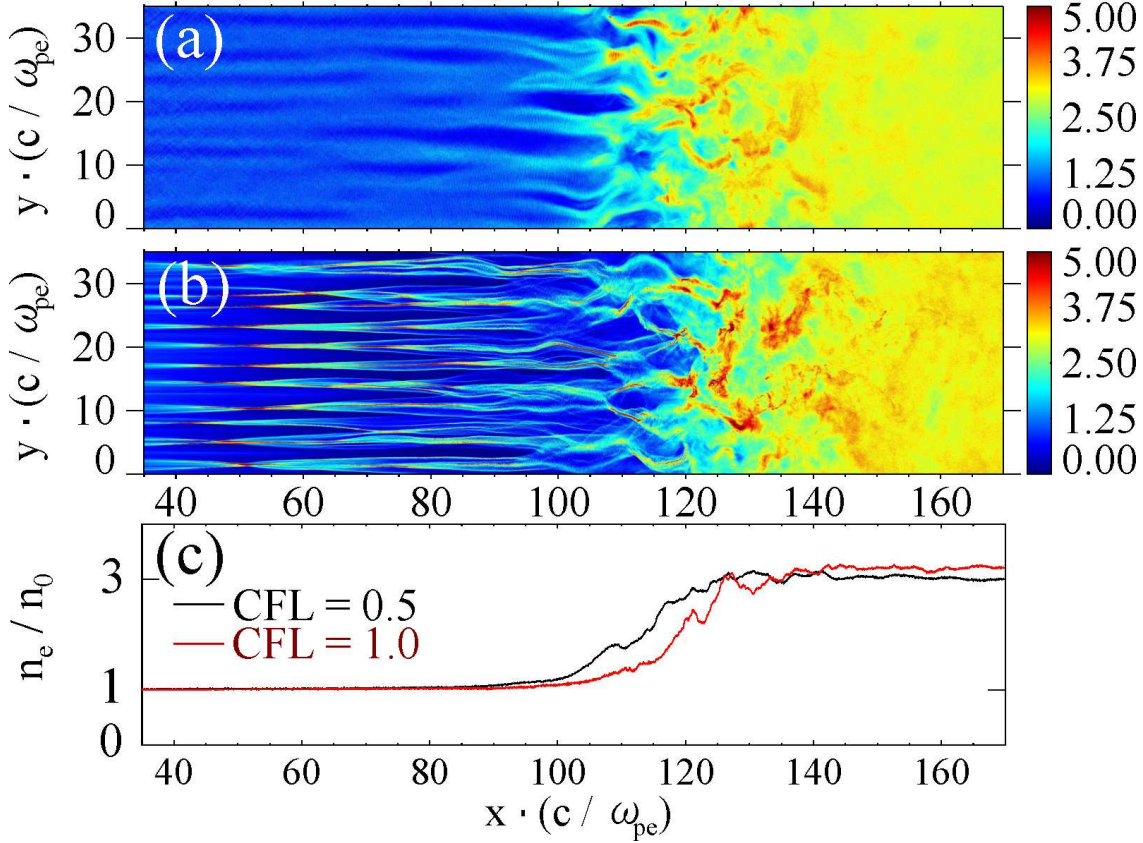}
    \end{center}
\caption{Electron density profile at $\omega_{pe} \cdot t = 177$ for CFL numbers of (a) 0.5 and (b) 1.0. (c) Transversely averaged electron density profile in the $x$-direction for $\sigma$ = 0.5 (black) and $\sigma$ = 1.0 (red). All quantities are normalized by the upstream value.}\label{fig:shock_den}
\end{figure}

\begin{figure}
 \begin{center}
   \FigureFile(160mm,200mm){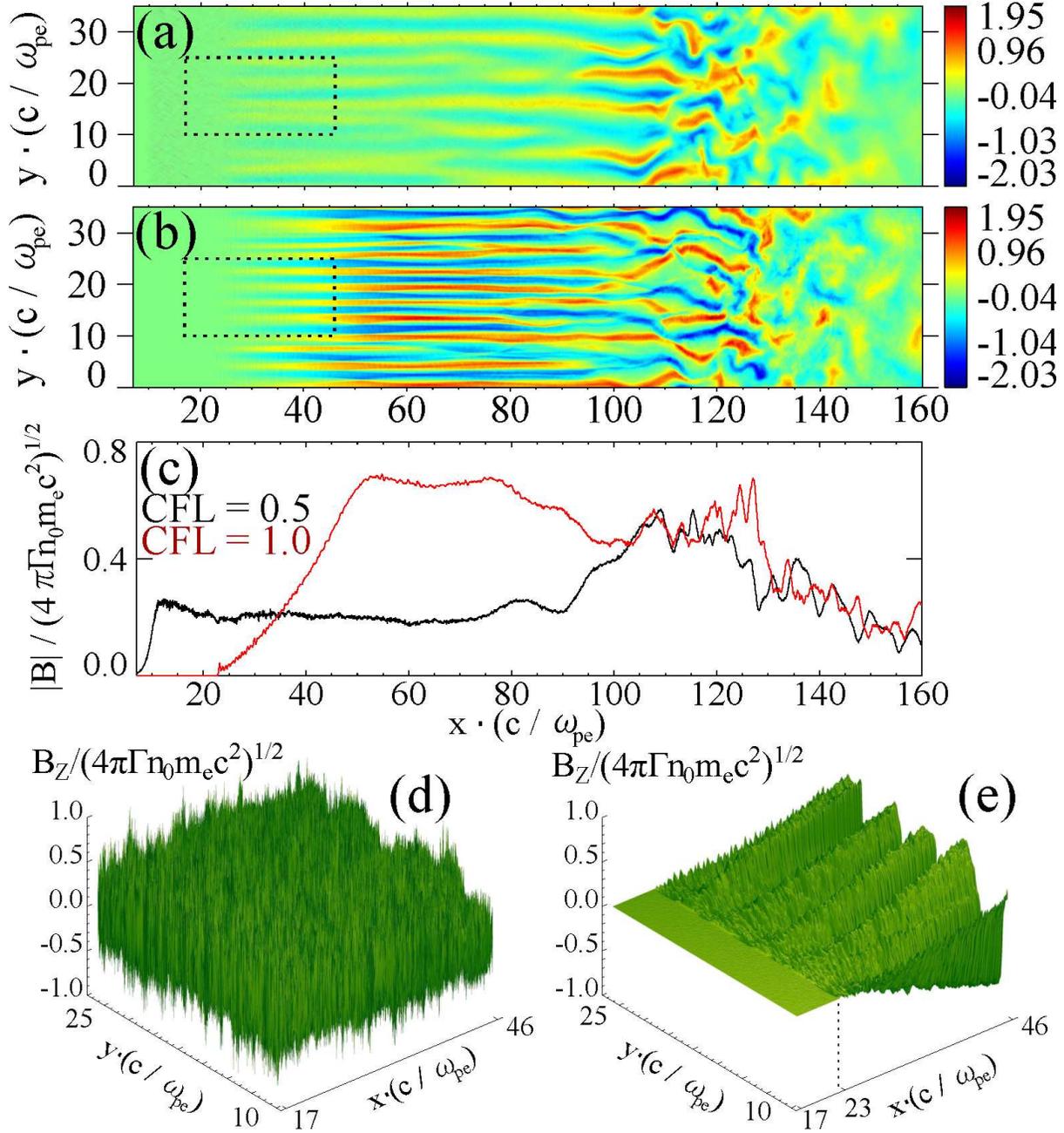}
 \end{center}
\caption{Spatial structure of the z-component of the magnetic field at $\omega_{pe} \cdot t$ = 177 for simulation runs with CFL numbers of (a) 0.5 and (b) 1.0. (c) Transversely averaged magnetic field profile in the $x$-direction for CFL = 0.5 (black) and CFL = 1.0 (red). (d) and (e) Enlarged view of the magnetic field profile in the area bordered by the dashed line in (a) and (b), respectively. All quantities are normalized by the upstream value.}\label{fig:shock_bz}

\end{figure}

Figure \ref{fig:shock_den} shows spatial structures of the electron number density at $\omega_{\mathrm{pe}} t$ = 177 after the shock front propagated upstream well away from the reflection wall. We compare simulation runs with different CFL numbers of 0.5 (Figure \ref{fig:shock_den} (a)) and 1.0 (Figure \ref{fig:shock_den} (b)). Although similar overall shock structures (e.g., position of the shock front and the jump condition in Figure \ref{fig:shock_den} (c)) were obtained, we found a difference in the filamentary density structure in the upstream between the two cases. The filaments clearly appeared when $\sigma$ = 1.0 but were blurred when $\sigma$ = 0.5.

There is a clear difference in the $z$ component of the magnetic field in Figure \ref{fig:shock_bz}. A high level of numerical noise ($\sim$ 10 \% of the upstream kinetic energy) was excited in the upstream as a result of the rapid growth of the numerical Cherenkov instability when $\sigma$ = 0.5 (Figure \ref{fig:shock_bz} (a) and Figure \ref{fig:shock_bz} (d)). This nonphysical generation of the magnetic field resulted in saturation of the Weibel instability with lower amplitudes (Figure \ref{fig:shock_bz} (a) and the black line in Figure \ref{fig:shock_bz} (c)). By contrast, in the case of $\sigma$ = 1.0, the Weibel instability had a clear filamentary structure (Figure \ref{fig:shock_bz} (b) and (e)), and the magnetic field energy generated by this instability reached 40\% of the upstream kinetic energy (red line in Figure \ref{fig:shock_bz} (c)).

\section{Summary}

We examined the stability property of the numerical Cherenkov instability using the PIC simulation package pCANS, which employs momentum-conserving field interpolation, the density decomposition method for the current deposit, and the implicit FDTD field solver for the Maxwell equations. We found that the numerical Cherenkov instability was remarkably inhibited with a CFL number of 1.0. This magical CFL number for the implicit FDTD method is larger than those found for explicit field solvers 0.5--0.7, which benefits the examination of long-term evolutions.

In addition, we showed another type of numerical Cherenkov instability that grew slowly in low wave number regions. This type of the instability is likely to arise from aliasing errors of the entropy mode carried by relativistic plasma flows, although could not simply understand from intersections between the electromagnetic mode and the beam aliases \citep{Godfrey_Vay_13}. This slowly growing mode can be suppressed with the adoption of higher-order shape functions by reducing the aliasing errors inherent in the PIC algorithm. We also found that the careful choice of the implicitness factor in the implicit field solver greatly reduced the growth rate. The optimal value for the second-order shape function was $\theta$ = 0.501, which has little impact on physical waves of interest in low wave number regions.

We followed evolutions of relativistic shocks by adopting the optimal parameters of the CFL number and the implicitness factor, and the second-order shape function. A shock wave formed clearly via the Weibel instability without growth of the numerical Cherenkov instability. The present stability properties of the numerical Cherenkov instability within the implicit FDTD scheme will allow us to explore long-term evolutions of multidimensional relativistic shock structures without generating nonphysical waves, which is crucial in the investigation of mechanisms of particle accelerations in collisionless shocks.
   
\bigskip

We would like to thank T. N. Kato and T. Hanawa for helpful discussions.


\begin{thebibliography}{}
\bibitem[Birdsall \& Langdon, (1991)]{Birdsall} Birdsall, ~C. ~K., \& Langdon, ~A. ~B. 1991, in Plasma Physics via Computer Simulation, ed.\ ~E. ~W. Laing (Bristol: IOP), ch.8, 164  
\bibitem[Dawson, (1983)]{Dawson} Dawson, ~J. ~M. 1983, Rev. Modern. Phys, 55, 403
\bibitem[Esirkepov, (2001)]{Esirkepov_01} Esirkepov, ~T. ~Zh. 2001, Comput. Phys. Commun, 135, 144--153
\bibitem[Godfrey, (1974)]{Godfrey_74} Godfrey, ~B. 1974, J. Comp. Phys, 15, 504--521
\bibitem[Godfrey \& Vay, (2013)]{Godfrey_Vay_13} Godfrey, ~B., \& Vay, ~J. ~L. 2013, J. Comp. Phys, 248, 33--46
\bibitem[Greenwood et al., (2004)]{Greenwood} Greenwood, ~A. ~D., Cartwright, ~K. ~L., Luginsland, ~J. ~W., Baca, ~E. ~A. 2004, J. Comp. Phys, 201, 665--684
\bibitem[Haber, (1973)]{Haber} Haber, ~I., Lee, ~R., Klein, ~H., Boris, ~J. 1973, Advance in electromagnetic simulation techniques, in: Proc, Sixth Conf. on Num, Sim. Plasmas, Berkeley, CA, 46--48 
\bibitem[Hoshino, (2013)]{Hoshino_13} Hoshino, ~M. 2013, \apj, 773, 118
\bibitem[Kato, (2007)]{Kato_07} Kato, ~T. ~N. 2007, \apj, 668, 974--979 
\bibitem[Lin et al., (1974)]{Lin} Lin, ~A. ~T., Dawson, ~J. ~M., Okuda, ~H. 1974, Phys. Fluids, 17, 1995 
\bibitem[Martins et al., (2010)]{Martins} Martins, ~S. ~F., Fonseca, ~R. ~A., Silva, ~L. ~O., Lu, W., Mori, ~W. ~B. 2010, Comput. Phys. Commun, 181, 869--875
\bibitem[Matsumoto et al., (2013)]{Matsumoto_13} Matsumoto, ~Y., Amano, ~T., Hoshino, ~M. 2013, \prl, 111, 215003
\bibitem[Nagata, (2008)]{Nagata} Nagata, ~K. 2008, in Interaction between Alternating Magnetic Fields and a Relativistic Collisionless Shock, phD thesis, Osaka University, ch2
\bibitem[Spitkovsky, (2008)]{Spitkovsky_08b} Spitkovsky, ~A. 2008, \apj, 682, L5--L8
\bibitem[Sironi \& Spitkovsky, (2011)]{Sironi_Spitkovsky_11} Sironi, ~L., \& Spitkovsky, ~A. 2011, \apj, 726, 75 
\bibitem[Vay et al., (2011)]{Vay_11} Vay, ~J. ~L., Geddes ~C. ~G. ~R., Cormier-Michel ~E., Grote ~D. ~P. 2011, J. Comp. Phys, 230, 5908--5929
\bibitem[Vay et al., (2013)]{Vay_13} Vay, ~J. ~L., Haber, ~I., Godfrey, ~B. 2013, J. Comp. Phys, 243, 260--268  
\bibitem[Xu et al., (2013)]{Xu_13} Xu, ~X., et al. 2013, Comput. Phys. Commun, 184, 2503--2514
\bibitem[Yu et al., (2014)]{Yu} Yu, ~P., Xu, ~X., Decyk, ~V. ~K., An, ~W., Vieira, ~J., Tsung, ~F. ~S., Fonseca, ~R. A., Lu, ~W., Silva, ~L. ~O., Mori, ~W. ~B. 2014, J. Comp. Phys, 266, 124--138
\end{thebibliography}
\end{document}